\documentclass[amsmath,amssymb,12pt]{revtex4}

\usepackage[final]{graphicx}
\DeclareGraphicsExtensions{.eps,.ps,.pdf}
\usepackage{amsmath,amssymb,bbold,bm}

\newcommand\vex[1]{\mathbf{#1}}
\newcommand\gvex[1]{\boldsymbol{#1}}

\def\d{\mathrm{d}}
\def\id{\mathbb{1}}
\def\tr{\mathrm{tr}}
\def\ii{\mathrm{i}}
\def\ee{\mathrm{e}}

\def\sslash#1{\setbox0=\hbox{$#1$}			
   \dimen0=\wd0                                		
   \setbox1=\hbox{/} \dimen1=\wd1  	 		
   \ifdim\dimen0>\dimen1                               	
      \rlap{\hbox to \dimen0{\hfil / \hfil}} 	  	
      #1                                      			
   \else                                        			
      \rlap{\hbox to \dimen1{\hfil$#1$\hfil}}   		
      \hbox{/} 	                              			
   \fi}   
\def\slash#1{\hbox{$#1$\kern-0.35em\raise0.1ex\hbox{/}}}

\def\prl{Phys. Rev. Lett. }
\def\prb{Phys. Rev. B }
\def\prd{Phys. Rev. D }

\begin{document}

\title{Midgap spectrum of the fermion-vortex system}
\author{B. Seradjeh}
\affiliation{Department of Physics and Astronomy, University of British Columbia,
Vancouver, BC, Canada V6T 1Z1}
\date{\today}

\begin{abstract}
I study the midgap spectrum of the fermion-vortex system in two spatial dimensions. The existence of bound states, in addition to the zero modes found by Jackiw and Rossi, is established. For a singly quantized vortex, I present complete analytical solutions in terms of generalized Laguerre polynomials in the opposite limits of vanishing and large vortex core size. There is an infinite number of such bound states, with a spectrum that is, when squared, given by, respectively, the Coulomb potential and the isotropic harmonic oscillator. Possible experimental signatures of this spectrum in condensed-matter realizations of the system are pointed out.
\end{abstract}

\maketitle

\section{Introduction}

The important role of topology in determining the properties of physical systems, whether in cosmology, particle physics, or condensed matter physics is by now well appreciated. This role is two-fold. On the one hand, topologically non-trivial structures or excitations such as instantons and vortices often arise in various physical systems, from quantum fields of elementary particles to ordered states of matter, and induce  fractional quantum numbers relative to the smooth background (i.e. vacuum sector)~\cite{JacReb76a}. Therefore, an understanding of their properties is required in order to fully understand the systems in question. On the other hand, certain physical systems sustain a novel type of order---topological order---that stems from their topological properties~\cite{Wen04a}. In two spatial dimensions, these two roles are closely related by the fact that topological excitations acquire fractional exchange statistics~\cite{OshSen06a}. The canonical example of this interplay is provided by quantum Hall states with their fractional excitations~\cite{Lau83b} and a ground-state degeneracy that depends on the topology of the space in which electrons move~\cite{WenNiu90a,TaoWu84a}. Due to its global character, a non-trivial topology also offers a special sort of protection against local perturbations, which could be useful for coherent manipulation of quantum states needed for quantum computation~\cite{Kit03a}.

A system where such phenomena arise is an ordered state of matter in which fermions interact with a vortex in the order parameter. If the topology of the vortex allows an odd number of midgap states, the fermion charge becomes fractionalized. When the spectrum is particle-hole symmetric, an odd number of midgap states will have zero energy, the so-called zero modes, and the fractional charge is found to be $\frac12$. The number of such zero modes is a topological invariant determined by index theorem. An interesting question is whether other midgap, non-zero-energy states exist. Apart from its intellectual appeal, a positive answer can be useful in detecting such structures in systems where they can potentially arise.

In superconductors and Fermi superfluids the vortex is carried in the pairing gap function. The core states of a vortex in an s-wave superconductor were first calculated by Caroli, de Gennes, and Matricon~\cite{CarGenMat64a}, later extended to other limits~\cite{CarMat65a}, and studied in great detail by others. In chiral superfluids and superconductors, the midgap spectrum was derived by Kopnin and Saloma~\cite{KopSal91a} and by Volovik~\cite{Vol99a}, and the number of zero modes was related to the vorticity~\cite{Vol93b}. However, these zero modes are not exact: In a p-wave superconductor, there is only one exact zero mode for odd vorticity~\cite{GurRad07a}. Moreover, since charge is not a good quantum number in a superconductor, the arguments for fractionalization of charge do not apply here. Nevertheless, the midgap spectrum has important consequences for thermodynamics and transport properties of vortices.

I study the midgap spectrum of a specific fermion-vortex system, where fermion number is a good quantum number and time-reversal symmetry is preserved, unlike quantum Hall systems where it is manifestly broken by the external magnetic field. It was originally studied by Jackiw and Rossi~\cite{JacRos81a}, who showed it supports a number, equal to vorticity, of \emph{exact} zero modes, unlike superconducting vortices. Weinberg~\cite{Wei81a} showed this number is protected by an index theorem. It was also studied~\cite{CugFraSch89a} in a lattice formulation, where similar zero modes were obtained. Recently, the same system has been shown~\cite{HouChaMud07a,SerWeeFra08a} to emerge in the low-energy sector of (spinless) fermions hopping on either a honeycomb or a $\pi$-flux square lattice, at half-filling, on the background of a vortex in the dimerized pattern of hopping amplitudes. Numerical study of the full lattice Hamiltonian shows~\cite{SerWeeFra08a} the number of exact zero modes and the fractional charge bound to the vortex survive beyond the low-energy theory. These vortices have been shown to obey fractional statistics~\cite{SerFra07a}. Similar proposals have also been advanced~\cite{Her07a,GhaWil07a,SerWebFra08a} to realize the zero modes in vortices of other possible types of order in graphene-based systems. I note that in Refs.~\cite{Her07a,SerWebFra08a} the zero modes are doubled by the valley index of a graphene layer. Since the vortex inevitably mixes the valleys (which are separated by a finite lattice momentum in the Brillouin zone) there is an intrinsic splitting of these doubled zero modes and they are not exact~\cite{SerWebFra08a}. The doubling also makes the number of zero modes always even so the fermion number is not fractional. Zero modes of Ref.~\onlinecite{GhaWil07a} occur in vortices of the superconducting order parameter. They are doubled by the spin degree of freedom and split by the Zeeman coupling.

In this paper, I derive the full midgap spectrum of the fermion-vortex system for a singly quantized vortex in two physically important limits---those of vanishing and large vortex core. Since the splitting of doubled zero modes could be treated as a perturbation, the midgap spectrum derived in this paper is relevant in such cases as well. The paper is organized as follows. In Sec.~\ref{sec:sys}, I describe the Hamiltonian. In Sec.~\ref{sec:spec}, I derive the equations, analyze the general asymptotic solutions, and present the complete midgap spectrum in the aforementioned limits. I conclude in Sec.~\ref{sec:concl} by examining the relevance of the two limits and with brief remarks on the relation to experiment. Some details of the solutions are given in an Appendix.

\section{Fermion-vortex system}\label{sec:sys}

The Hamiltonian is given by
\begin{equation}\label{eq:ham}
H=\vex p\cdot\gvex\alpha+|m|\beta \ee^{\ii\chi\gamma_5},
\end{equation}
where $\vex p=-\ii\gvex\nabla$ is the momentum operator and the matrices $\gvex\alpha=(\gamma_0\gamma_1,\gamma_0\gamma_2)$, $\beta=\gamma_0$ furnish a four dimensional Clifford algebra. For concreteness I use the Weyl representation in which $\gamma_0=\sigma_1\otimes\id$, $\gamma_k=-\ii\sigma_2\otimes\sigma_k$ ($k=1,2,3$) and $\gamma_5=\ii\gamma_0\gamma_1\gamma_2\gamma_3=\sigma_3\otimes\id$ with $\sigma_k$ the Pauli matrices. The scalar mass $m=|m|\ee^{\ii\chi}$ contains a $v$-vortex located at the origin, i.e., the phase $\chi$ winds by $2\pi v$ around a loop containing the origin, the amplitude $|m|$ takes a constant value, $m_\infty$, at infinity and vanishes as $m_0 r^v$ at the origin. We take a radially symmetric form, $m(r,\theta)=|m(r)|\ee^{\ii v\theta}$, in polar coordinates $(r,\theta)$.

\section{Midgap spectrum}\label{sec:spec}

We wish to solve the bound-state spectrum of~(\ref{eq:ham}),
\begin{equation}\label{eq:eigeq}
H\psi = E\psi,\quad |E|<m_\infty.
\end{equation}
The eigenstate is written as $\psi=U(\psi_a,\psi_b)^T$ where
$$
U=\left[ \begin{array}{cc} P_+ & P_- \\ P_- & P_+ \end{array} \right]
$$
is unitary and the projections $P_\pm=\frac12(1\pm\sigma_3)$. In this basis, the eignevalue equation~(\ref{eq:eigeq}) is reduced to
\begin{subequations}
\begin{eqnarray}
h\psi_b = E\psi_a, &&\quad
h^\dag\psi_a = E\psi_b, \\
h 
&=&  \left[ \begin{array}{cc} m^* & \pi \\ -\pi^\dag & m \end{array} \right],
\end{eqnarray}
\end{subequations}
with $\pi\equiv p_x+\ii p_y$. We may solve these equations by solving one of two squared equations
\begin{equation}
h^{}h^\dag\psi_a=E^2\psi_a, \quad \psi_b=\frac1E h^\dag\psi_a; \label{eq:sqra}
\end{equation}
or
\begin{equation}
\quad h^\dag h^{}\psi_b=E^2\psi_b, \quad \psi_a=\frac1E h^{}\psi_b. \label{eq:sqrb}
\end{equation}
We note that $\pi\pi^\dag=\pi^\dag\pi=\vex p^2$, and 
\begin{equation}
hh^\dag=\left[ \begin{array}{cc} |m|^2+\pi\pi^\dag & [\pi,m^*] \\ {[m,\pi^\dag]} & |m|^2+\pi^\dag\pi \end{array} \right].
\end{equation}

For a $v$-vortex,
\begin{equation}
[\pi, m^*] = -\ii \ee^{\ii(1-v)\theta}\left(\frac{\d|m|}{\d r}+\frac{v|m|}r\right).
\end{equation}
For $v>0$, I shall solve the set of Eqs.~(\ref{eq:sqra}), which includes the Jackiw-Rossi zero modes with $\psi_b=0$. Similarly, for $v<0$ we need to solve Eqs.~(\ref{eq:sqrb}). Since $\tr(\sigma_3 h^{}h^{\dag})=0$ we may take $\psi_a=(f,\eta f^*)^T$, $\eta=\pm1$,
\begin{equation}
\left(\vex p^2+|m|^2-E^2\right) f - \ii\eta\ee^{\ii(1-v)\theta}\left(\frac{\d|m|}{\d r}+\frac{v|m|}r\right) f^* = 0.
\end{equation}
The dependence on $\theta$ can be eliminated using a two-phase ansatz,
\begin{equation}\label{eq:2phase}
f(r,\theta)=\ee^{-\ii\ell_1\theta}f_1(r)+\ee^{-\ii\ell_2\theta}f_2(r), \quad \ell_1+\ell_2=v-1.
\end{equation}
Finally we remove the complex phases by writing $f_{i}=\ee^{\ii\alpha_{i}}\phi_{i}$, $i=1,2$ such that $\ii\ee^{\ii(\alpha_1+\alpha_2)}=\pm1\equiv\zeta$. Thus,
\begin{subequations}\label{eq:gen}\begin{eqnarray}
\left(-\frac1r\frac{\d}{\d r}r\frac{\d}{\d r}+\frac{\ell_1^2}{r^2} + |m|^2-E^2\right)\phi_1 + \zeta\eta\left(\frac{\d|m|}{\d r}+\frac{v|m|}r \right) \phi_2 = 0, \\
\left(-\frac1r\frac{\d}{\d r}r\frac{\d}{\d r}+\frac{\ell_2^2}{r^2} + |m|^2-E^2\right)\phi_2 + \zeta\eta\left(\frac{\d|m|}{\d r}+\frac{v|m|}r \right) \phi_1 = 0.
\end{eqnarray}\end{subequations}

At $r\to\infty$, Eqs.~(\ref{eq:gen}) are solved by $\phi_i(r)=C_i\ee^{-\sqrt{m_\infty^2-E^2}r}$. For $r\to 0$ we have to the leading order, $\phi_i=D_i r^{a_i}$ with
\begin{subequations}\begin{eqnarray}
(\ell_1^2-a_1^2)D_1 r^{a_1} + 2\zeta\eta (v+1)m_0 D_2 r^{a_2+v+1} = 0, \\
(\ell_2^2-a_2^2)D_2 r^{a_2} + 2\zeta\eta (v+1)m_0 D_1 r^{a_1+v+1} = 0,
\end{eqnarray}\end{subequations}
which can be solved (to leading order in $r$) if either $a_1=|\ell_1|,\ a_2=a_1+v+1$, or $a_2=|\ell_2|,\ a_1=a_2+v+1$. In both cases $a_1,a_2>0$ guarantees the regularity at the origin. The condition for having a solution over the entire range of $r$ is the smooth matching of the two asymptotic solutions. For this we need both asymptotic solutions at the origin. This matching would also determine the energy levels. 

I present explicit solutions for the case $\ell_1^2=\ell_2^2\equiv\ell^2$. For $v=1$, we have $\ell_1+\ell_2=0$, so this condition is satisfied by all the solutions. For $v>1$, on the other hand, this means $\ell_1=\ell_2=\frac{v-1}2$, and we find just a small subset of solutions. 

Then, Eqs.~(\ref{eq:gen}) collapse into one, $\phi_1=-\zeta\eta\phi_2\equiv\phi,$ and take the form of the radial part of Schr\"odinger equation in two dimensions
\begin{equation}\label{eq:red}
\left(-\frac1r\frac{\d}{\d r}r\frac{\d}{\d r}+\frac{\ell^2}{r^2} + V(r) -E^2\right)\phi = 0,
\end{equation}
with an effective potential
\begin{equation}\label{eq:eff}
V(r) = |m|^2  - \frac{v|m|}r - \frac{\d|m|}{\d r}.
\end{equation}
The sign $-\zeta\eta$ is chosen to allow for bound state solutions to exist. 

When $v=1$ we may relabel the solutions with $\ell$ as the angular momentum. From Eq.~(\ref{eq:2phase}) we find $f_\ell=(\ee^{-\ii\ell\theta}\ee^{\ii\alpha_1}+\ii\eta\ee^{-\ii\alpha_1}\ee^{\ii\ell\theta})\phi$. By taking linear combinations, $(\psi_a)_\ell\pm\ii\eta\ee^{\pm2\ii\alpha_1}(\psi_a)_{-\ell}$, and suppressing overall constant factors the final solutions can be labeled by an angular momentum $\ell$ as 
\begin{equation}\label{eq:psia}
\psi_a=\ee^{\ii\ell\theta}\phi \left[\begin{array}{c}1 \\ -\ii \end{array}\right].
\end{equation}

I now examine the two opposite limits where we can completely solve the spectrum for $v=1$.

\subsection{Quantum limit}

When the core size of the vortex is vanishingly small, we can take $|m|$ to be a constant, $m_\infty$. Then Eq.~(\ref{eq:eff}) is reduced to the two-dimensional Coulomb potential, $V^>(r)=m^2_\infty-vm_\infty/r$. Taking $v=1$, I define $\phi(r)=r^{\ell}\ee^{-\mu r}\varphi(r)$, with $\mu=\sqrt{m_\infty^2-E^2}$, and redefine the variable $z\equiv2\mu r$ to find
\begin{equation}\label{eq:L}
\frac{\d^2}{\d z^2}\varphi+\left(\frac{2\ell+1}z-1 \right)\frac{\d}{\d z}\varphi - \left(\ell+\frac12 -\frac{m_\infty}{2\mu} \right)\varphi=0.
\end{equation}
The solutions to Eq.~(\ref{eq:L}) are given in terms of generalized Laguerre polynomials, $L_s^{(\nu)}(z)$, where $\nu=2\ell$ and $s=\frac{m_\infty}{2\mu}-\ell-\frac12\in\mathbb{N}\cup\{0\}$. So, the spectrum is given by
\begin{equation}\label{eq:qlspec}
\phi^>_{n,\ell}(r) = r^\ell \ee^{-\frac{m_\infty}nr}L_{\frac{n-1}2-\ell}^{(2\ell)}\left(\frac{2m_\infty}n r\right),\quad E^>_n = \pm m_\infty\sqrt{1-\frac1{n^2}},\quad n\in \mathbb{N},
\end{equation}
where $n=2(s+\ell)+1\in\mathbb{N}$. 

Since $s\geq0$, the maximum value $\ell$ can assume for fixed $n$ is found when $s=0$, i.e. $\ell\leq\frac{n-1}2$. The integrability of $|\phi^>_{n,\ell}|^2$ at the origin restricts the minimum value of $\ell \geq -\frac{n-1}2$. To see this, first note that at the origin $r|\phi^>_{n,\ell}(r)|^2\sim r^q$ with $q=2\ell+1+2p$, where $p$ is the smallest power of $L_s^{(\nu)}$. Integrability requires $q>-1$. For $\nu<0$, $p=-\nu\Theta(\nu+s)$, where $\Theta$ is the Heaviside step function. Thus, $q+1=\ell+1-2\ell\Theta(2\ell+s)>0$. Since $\ell+1\leq0$, we must have $2\ell+s\geq0$, i.e. $\ell\geq-\frac{n-1}2$. Therefore, $|\ell|\leq\frac{n-1}2$ and the degeneracy of $E_n$ is $g_n=n$. 

These values of $\ell$ also enforce the boundary condition $f_{n,\ell}(r,\theta+2\pi)=(-1)^{n+1}f_{n,\ell}(r,\theta)$, or conversely, the given boundary condition selects odd or even values of $n$. The complete normalized solutions (including $\psi_b$) are given in the Appendix.

When $v>1$, we find a subset of solutions with $2\ell=v-1$, $n=2s+v\geq v$ by increments of 2, and $E^>_n=\pm m_\infty\sqrt{1-\frac{v^2}{n^2}}$. In particular, we find only one of the $v$ zero modes, but we also find one non-zero mode per each orbital $n$.


\subsection{Core limit}

Deep inside the vortex core we may take the scalar mass to be $|m(r)|=m_0 r^v$ everywhere. Then the effective potential, Eq.~(\ref{eq:eff}), reads
\begin{equation}
V^<(r)=A r^{2v} - B r^{v-1},
\end{equation}
with $A=m_0^2$ and $B=2vm_0$. The spectrum of this potential for general $v$ and $E\neq0$ is, to the best of my knowledge, not known in closed form. For $E=0$ closed-form bound state solutions are found~\cite{MayVin92a}, when 
\begin{equation}\label{eq:cond}
\frac{B}{2(v+1)\sqrt{A}}-\frac12-\frac{|\ell|}{v+1}=k \in\mathbb{N}\cup\{0\},
\end{equation}
again in terms of generalized Laguerre polynomials, 
\begin{equation}
\phi^<_{k,\ell}=r^{|\ell|}\ee^{-\frac{m_0}{v+1}r^{v+1}}L_k^{\left(\frac{2|\ell|}{v+1}\right)}\left(\frac{2m_0}{v+1}r^{v+1}\right).
\end{equation}
The bound-state condition, in our case, is $v-1-2|\ell|=2(v+1)k$, which is satisfied only for $k=0$. Since $L_0^{(\alpha)}=1$, this just confirms the explicit solution found originally by Jackiw and Rossi~\cite{JacRos81a}. In contrast to the quantum limit, the non-zero modes are not found for $v>1$ in this limit.

When $v=1$, the potential $V(r)=m_0^2 r^2-2m_0$ is that of an isotropic harmonic oscillator. So, now the solutions for non-zero $E$ can be found, by shifting $B\to B+E^2$ in Eq.~(\ref{eq:cond}), as
\begin{equation}\label{eq:clspec}
\phi^<_{n,\ell} = r^{|\ell|}\ee^{-\frac12m_0r^2}L_{\frac{n-1-|\ell|}2}^{(|\ell|)}(m_0r^2), \quad E^<_n = \pm\sqrt{2m_0(n-1)}, \quad n\in \mathbb{N},
\end{equation}
where $|\ell| \leq n-1$ by increments of 2. Thus, the degeneracy of $E_n$ is again $g_n=n$. Since $\ell\in\mathbb{Z}$, the boundary condition is necessarily $f_{n,\ell}(r,\theta+2\pi)=f_{n,\ell}(r,\theta)$, unlike the quantum limit where antiperiodic boundary conditions are also accommodated.

\section{Conclusion}\label{sec:concl}

In sum, I presented, in two physically interesting limits, analytical solutions for the midgap spectrum of the vortex-fermion system whose zero modes were found by Jackiw and Rossi~\cite{JacRos81a}. This system has recently emerged as the low-energy theory of certain condensed matter systems, in particular in some ordered states of graphene-based structures. I will briefly discuss the relevance of these two limits and remark on possible experimental signatures of the midgap spectrum in condensed-matter realizations of the Hamiltonian considered here.

The core size, $\xi$, of the vortex can be roughly estimated as $\xi\simeq m_\infty/m_0$. The quantum and core limits are found, respectively, in the region where $r\gg \xi$ and $r\ll \xi$. Since the zero mode is topologically protected, it is found in both limits. The ``high-energy'' part, where $E\lesssim m_\infty$, is dominated by the quantum-limit spectrum, Eq.~(\ref{eq:qlspec}), while the non-zero ``low-energy'' part of the midgap spectrum, where $E\ll m_\infty$, is dominated by the core-limit spectrum, Eq.~(\ref{eq:clspec}). The low-energy spectrum merges with the high-energy one when the orbital quantum number, $n=n_c\simeq\xi m_\infty$. This implies the low-energy part exists when $n_c\gg1$, i.e. $\sqrt{m_0}\ll m_\infty$.

For small enough $\frac{\sqrt{m_0}}{m_\infty}$ the low-energy spectrum can be treated as continuous with a density of states 
$$
\rho^<(E)=\sum_n g_n\delta(E-E^<_n)=\frac{E^3}{(2m_0)^2}.
$$
By contrast, the density of states of the high-energy spectrum is 
$$
\rho^>(E)=\sum_n g_n\delta(E-E^>_n)=\frac{m_\infty^2E}{(m_\infty^2-E^2)^2},
$$
which has a sharp divergence at $E=m_\infty$. This spectrum must be observable in STM experiments that probe the local density of states. It must also have thermodynamic signatures. For instance, at low temperatures $\sqrt{m_0}\ll T\ll m_\infty$ the low-energy spectrum renders the specific heat $C_V\sim T^4$. Although to detect such thermodynamic dependencies experimentally a large enough volume of the system must be occupied by the vortex matter.

\section*{Acknowledgment}
The author has benefited from discussions with E. Fradkin, M. Franz and I. Herbut. This research has been supported by NSERC, CIfAR and the Killam Foundation. 

\appendix*
\section{Normalized solutions in quantum limit}

Using Eq.~(\ref{eq:sqra}) and Eq.~(\ref{eq:psia}) I write
\begin{equation}
\psi_b=\frac1E\ee^{\ii\ell\theta} \left[\begin{array}{c} \ee^{\ii\theta}\left( \frac{\d}{\d r} - \frac{\ell}r + m_\infty\right)\phi \\ -\ii\ee^{-\ii\theta}\left( \frac{\d}{\d r} + \frac{\ell}r + m_\infty\right)\phi \end{array}\right].
\end{equation}

Using the properties of generalized Laguerre functions, namely
\begin{equation}
\frac{\d}{\d z}L_s^{(\nu)}(z) = - L_{s-1}^{(\nu+1)}(z),
\end{equation}
and the recurrence relation
\begin{equation}\label{eq:recur}
L_s^{(\nu)}(z) = L_s^{(\nu+1)}(z) - L_{s-1}^{(\nu+1)}(z),
\end{equation}
from Eq.~(\ref{eq:qlspec}) we find for level $n$ and $\ell$,
\begin{equation}
\left(\psi_b^>\right)_{n,\ell} = \pm\frac1{\mathcal{N}_{n,\ell}^>}\frac{n}{\sqrt{n^2-1}} \ee^{\ii\ell\theta}r^\ell\ee^{-\frac{m_\infty}n r}
\left[\begin{array}{c} \ee^{\ii\theta}\left(L_s^{(\nu)}(z)-\frac{L_s^{(\nu+1)}(z)+L_{s-1}^{(\nu+1)}(z)}n\right) \\
-\ii \ee^{-\ii\theta}\left(\left[1+\frac{\nu}{m_\infty r}\right]L_s^{(\nu)}(z)-\frac{L_s^{(\nu+1)}(z)+L_{s-1}^{(\nu+1)}(z)}n\right) \end{array}\right],
\end{equation}
where $\mathcal{N}_{n,\ell}^>$ ensures proper normalization, and I remind the reader that $\nu=2\ell$, $s=\frac{n-1}2-\ell$, and $z=\frac{2m_\infty}n r$.

The normalization factor $\mathcal{N}_{n,\ell}^>$ can be found by first noting that from Eqs.~(\ref{eq:sqra})
\begin{equation}
|\psi_b|^2=\frac1{E^2}\psi_a^*h^{}h^\dag\psi_a=|\psi_a|^2.
\end{equation}
Using the integral
\begin{equation}
\int_0^\infty z^{\nu+1}\ee^{-z}\left[ L_s^{(\nu)}(z)\right]^2 \d z = \frac{(\nu+s)!}{s!}(2s+\nu+1),
\end{equation}
We can then find
\begin{equation}
\mathcal{N}_{n,\ell}^> = \sqrt{8\pi n}\left(\frac n{2m_\infty}\right)^{\ell+1} \sqrt{\frac{\left(\frac{n-1}2+\ell\right)!}{\left(\frac{n-1}2-\ell\right)!}}.
\end{equation}


\end{document}